\documentclass[conference]{IEEEtran}
\IEEEoverridecommandlockouts
\usepackage{cite}
\usepackage{amsmath,amssymb,amsfonts}
\usepackage{algorithmic}
\usepackage{graphicx}
\usepackage{textcomp}
\usepackage{colortbl}
\definecolor{G}{RGB}{144,238,144}
\definecolor{R}{RGB}{255,160,122}
\usepackage{xcolor}
\def\BibTeX{{\rm B\kern-.05em{\sc i\kern-.025em b}\kern-.08em
    T\kern-.1667em\lower.7ex\hbox{E}\kern-.125emX}}
\begin{document}

\title{Attentional Triple-Encoder Network in Spatiospectral Domains for Medical Image Segmentation
}
\author{
    \IEEEauthorblockN{Kristin Qi\textsuperscript{1}, Xinhan Di\textsuperscript{2} }\IEEEauthorblockA{\textsuperscript{1}University of Massachusetts Boston, USA \textsuperscript{2}Giant Network AI Lab, China \\yanankristin.qi001@umb.edu, dixinhan@ztgame.com}
    }

\maketitle
\begin{abstract}
Retinal Optical Coherence Tomography (OCT) segmentation is essential for diagnosing pathology. Traditional methods focus on either spatial or spectral domains, overlooking their combined dependencies. We propose a triple-encoder network that integrates CNNs for spatial features, Fast Fourier Convolution (FFC) for spectral features, and attention mechanisms to capture global relationships across both domains. Attention fusion modules integrate convolution and cross-attention to further enhance features. Our method achieves an average Dice score improvement from 0.855 to 0.864, outperforming prior work.  
\end{abstract}
 
\begin{IEEEkeywords}
OCT image segmentation,  attention mechanism
\end{IEEEkeywords}

\section{Introduction}


Optical Coherence Tomography (OCT) is an imaging tool in ophthalmology to analyze retinal structures. By providing high-resolution, cross-sectional images, OCT facilitates the diagnosis of pathologies. Automated segmentation enhances the utility of OCT by identifying intraretinal fluid regions and retinal layers, thereby improving diagnostic efficiency.

Traditional segmentation methods, including graph-based approaches and convolutional neural networks (CNNs), form the foundation for OCT segmentation. Recently, attention-based models with U-Net demonstrates improved feature extraction by effectively capturing spatial relationships and dependencies \cite{cao2023swin}. However, these methods often focus exclusively on either spatial or spectral domains, limiting their ability to handle both domains and their relationships \cite{farshad2022net}.

To address these limitations, we propose a triple-encoder network with dual-domain transformers to enhance spatial and frequency-domain features. This network integrates CNNs for spatial features, Fast Fourier Convolution (FFC) for spectral features, and employs attention mechanisms to capture global feature relationships across both domains. The network is structured into two stages.
 
\begin{figure}[t]
    \centering
    \includegraphics[width=\linewidth]{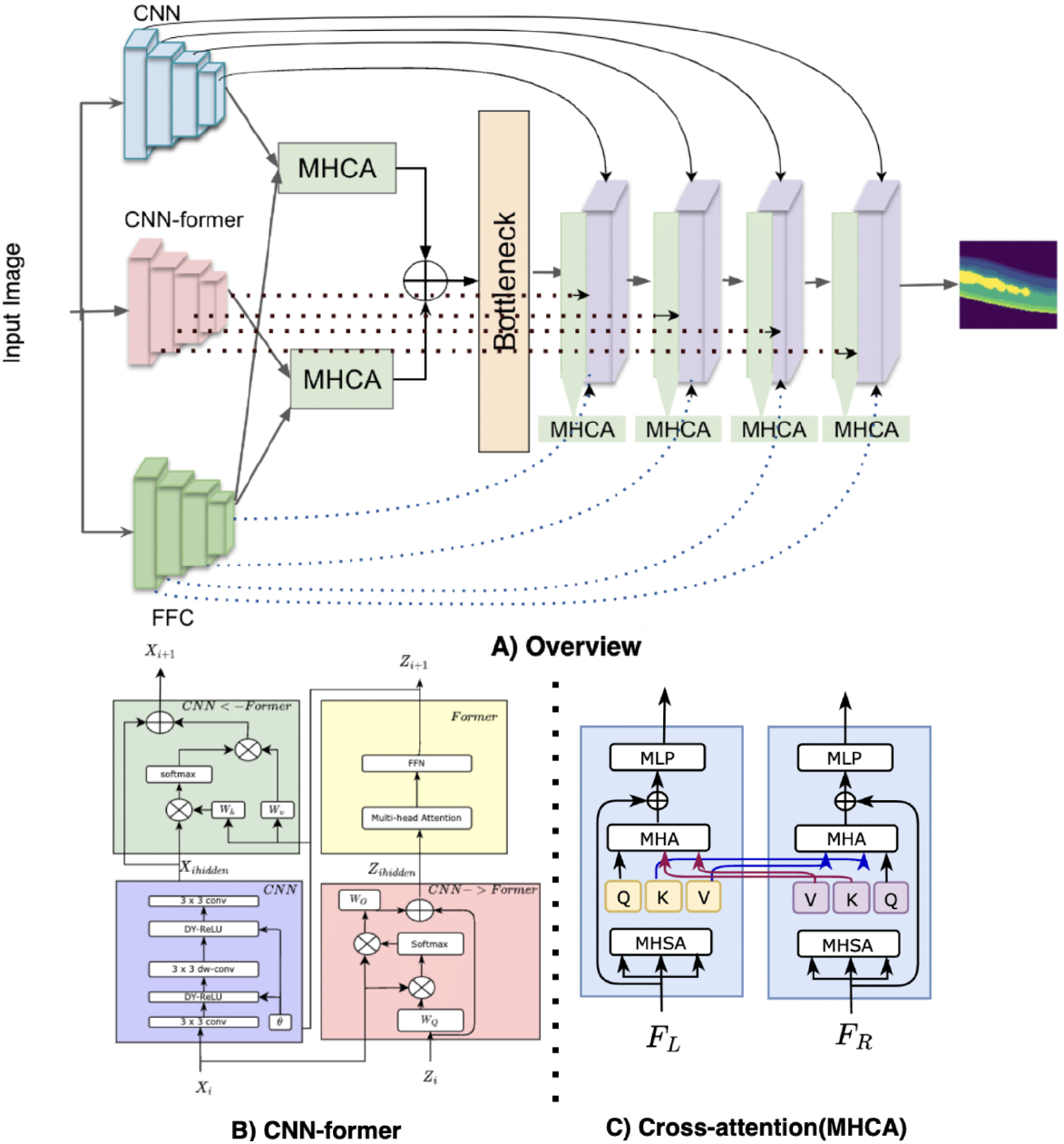}
    \caption{
        (A) Overview: multi-head cross-attention (MHCA) fuses features from three encoder branches. 
        (B) Details of the CNN-former, consisting of four sub-blocks for feature enhancement. 
        (C) Details of the MHCA, showing how bidirectional cross-attention processes two feature sets ($F_L$ and $F_R$).
    }
    \label{fig:network_pipeline}
\end{figure}

\textbf{Stage 1:} The encoders enhance spatial and spectral features by aggregating domain-specific information and increasing global context \cite{farshad2022net}. It ensures a representation that captures both local details and long-range dependency. \textbf{Stage 2:} Attention fusion modules integrate the spatial and spectral features from Stage 1. These modules use cross-attention to further improve features. Specifically, convolutional layers are combined with cross-attention both in the encoder and the bottleneck layer (see Figure \ref{fig:network_pipeline}). The decoder then integrates these features, leveraging the cross-attention for effective feature selection and convolution for capturing spatial details \cite{farshad2022net}. Compared to  prior work \cite{farshad2022net}, our network achieves a higher average Dice score for improved segmentation accuracy.

\section{Method}
The proposed network is illustrated in Figure 1-A. 1-B and 1-C are details of specific attention mechanisms used.

\textbf{Stage 1:} Three encoders improve features in both spatial and spectral domains. The spatial encoder processes local details, the spectral encoder handles frequency-domain information, and the attention encoder enhances global relationships. Cross-attention modules merge complementary features into a unified representation.

For an input image, three encoder branches—CNN, FFC, and CNN-former—are combined to process features. The CNN and FFC each have four blocks for spatial and spectral encoders, respectively. The CNN-former aligns features from both domains via combining attention and convolution to establish one-to-one correspondence in four sub-blocks shown below \cite{chen2022mobile}. The CNN-former also fuses FFC for unified feature alignment before further processing. 

\noindent\textbf{CNN sub-block:} Processes OCT image input using Dynamic ReLU and two multilayer perceptron (MLP) layers, producing feature maps for CNN $\leftarrow$ Former.

\noindent\textbf{Former sub-block:} Refines features with multi-head attention transformers and feed-forward networks.

\noindent\textbf{CNN $\rightarrow$ Former:} Applies lightweight cross-attention to map CNN features into attention tokens \cite{chen2022mobile}.

\noindent\textbf{CNN $\leftarrow$ Former:} Maps refined attention tokens back to CNN features using lightweight cross-attention.

\textbf{Stage 2:} This stage receives concatenated feature maps from Stage 1 and further enhances them using cross-attention and convolutional layers. Two cross-attention modules generate relational weights to integrate features from CNN, FFC, and CNN-former (see Equation 1), enhancing integrated features \cite{farshad2022net}. The decoding process iteratively fuses these modules across multiple levels using four cross-attention modules. This maintains contextual relationships across domains  and generates final segmentation maps. Two cross-attention modules (bidirectional) are defined as:
\[
\begin{aligned}
& \operatorname{Attn}=\operatorname{Attn}_{\text {cross }}\left(\tilde{f}^{cnn}, \tilde{f}^{ffc}\right)+\operatorname{Attn}_{\text {cross }}\left(\tilde{f}^{ffc}, \tilde{f}^{former}\right)   
\end{aligned}\tag{1}
\].

\noindent$\tilde{f}^{cnn}$ is feature map from the CNN in Stage 1. $\tilde{f}^{ffc}$ is feature map from the FFC. $\tilde{f}^{former}$ is feature map from CNN-former.

To optimize training, a combined loss function is used:
\[\mathcal{L}_{\text {total }}=\lambda_{\text {Dice }} \mathcal{L}_{\text {Dice }}+\lambda_{C E} \mathcal{L}_{C E}, \tag{2}
\] where $\mathcal{L}_{\text {Dice }}$ is Dice loss \cite{sudre2017generalised}. $\mathcal{L}_{C E}$ is cross-entropy loss.  $\lambda_{D i c e}$ and $\lambda_{C E}$ balance the two loss terms.




\section{Results}
We evaluated the proposed network on the Duke OCT dataset, which consists of scans from 10 patients. The dataset was divided into three subsets: training (subjects 1–6), validation (subjects 7–8), and testing (subjects 9–10). For evaluation, annotations provided by Expert-1 were the reference. The training process follows the parameters in prior work \cite{farshad2022net}, with images resized to 224 $\times$ 224, a learning rate of $5 \times 10^{-4}$, weight decay of $1 \times 10^{-4}$, Adam optimizer, and trained 100 epochs. The loss function balances Dice loss ($\lambda_{\text{Dice}} = 1$) and cross-entropy loss ($\lambda_{CE} = 1)$.

Table 1 shows that the proposed network achieves higher performance than most existing methods. The average Dice score improved from 0.855 to 0.864, indicating that the network effectively fuses spatial and spectral domain features. Particularly, the network achieves the Dice score of 0.94 for the fluid layer and 0.87 for the ILM layer, showing its effectiveness on challenging regions.


\begin{table}[t]
\label{fig:first}
\scriptsize
\begin{center}
\caption{Per-layer dice score and overall conparisons.}
\setlength\tabcolsep{2pt} 
\renewcommand{\arraystretch}{1.2} 
\begin{tabular}{@{\hskip 0pt}l@{\hskip 3pt}c@{\hskip 3pt}c@{\hskip 3pt}c@{\hskip 3pt}c@{\hskip 3pt}c@{\hskip 3pt}c@{\hskip 3pt}c@{\hskip 3pt}c@{\hskip 3pt}c@{\hskip 0pt}}

\hline
Method & ILM & NFL-IPL & INL & OPL & ONL-ISM & ISE & OS-RPE & Fluid & Mean \\
\hline
RelayNet \cite{roy2017relaynet} & $0.84$ & $0.85$ & $0.70$ & $0.71$ & $0.87$ & $0.88$ & $0.84$ & $0.30$ & $0.75$ \\
\hline
Language \cite{tran2020retinal} & $0.85$ & $0.89$ & $0.75$ & $0.75$ & $0.89$ & \cellcolor{G}$\mathbf{0 . 9 0}$ & \cellcolor{G}$\mathbf{0 . 8 7}$ & $0.39$ & $0.78$ \\
\hline
Alignment \cite{maier2021line} & $0.85$ & $0.89$ & $0.75$ & $0.74$ & \cellcolor{G}$\mathbf{0.90}$ & \cellcolor{G}$\mathbf{0.9 0}$ & \cellcolor{G}$\mathbf{0 . 8 7}$ & $0.56$ & $0.81$ \\
\hline
U-Net \cite{ronneberger2015u} & $0.84$ & $0.89$ & $0.77$ & $0.76$ & $0.89$ & $0.89$ & $0.85$ & $0.80$ & $0.836$ \\
\hline
Y-Net \cite{farshad2022net} &$0.86$ & $0.89$ & $0.78 $ & $0.75$ & \cellcolor{G}$\mathbf{0.90}$ & $0.88$ & $0.85$ & $0.93$ & $0.855$ \\
\hline
Ours  & \cellcolor{G}$\mathbf{0.87}$ & \cellcolor{G}$\mathbf{0.90}$ & \cellcolor{G}$\mathbf{0.79}$ & \cellcolor{G}$\mathbf{0.77}$ & \cellcolor{G}$\mathbf{0.90}$ & $0.89$ & $0.85$ & \cellcolor{G}$\mathbf{0.94}$ & \cellcolor{G}$\mathbf{0.864}$\\
\hline
\end{tabular}
\end{center}
\end{table}

\noindent\textbf{Ablation.} As shown in Table \ref{table:ablation}, we evaluated two settings: 1)
removing cross-attention in the Stage 1 encoder \textbf{(w/o)} and 2) removing cross-attention in the decoder \textbf{(w/o decoder)}. Both configurations result in reduced performance, indicating the effectiveness of cross-attention in fusing both spatial and spectral features for feature enhancement.

\begin{table}[!t]
\scriptsize 
\centering
\caption{Ablation study results for different cross-attention settings.}
\setlength\tabcolsep{2.5pt} 
\renewcommand{\arraystretch}{1.2} 
\begin{tabular}{lccccccccc}
\hline
Setting & ILM & NFL-IPL & INL & OPL & ONL-ISM & ISE & OS-RPE & Fluid & Mean \\
\hline
w/o & $0.87$ & $0.90$ & $0.77$ & $0.75$ & $0.90$ & $0.89$ & $0.86$ & $0.92$ & $0.858$ \\
w/o decoder & $0.86$ & $0.90$ & $0.78$ & $0.76$ & $0.90$ & $0.89$ & $0.85$ & $0.94$ & $0.86$ \\
\hline
\end{tabular}
\label{table:ablation}
\end{table}

\section{Conclusion}
We proposed a triple-encoder network with attention mechanisms to effectively fuse spatial and spectral features, enabling higher segmentation performance for OCT images. Future work will validate the network on other OCT datasets.

\vspace{-1.15pt} 
\bibliographystyle{IEEEtran}
\bibliography{mybib}

\end{document}